\documentclass[apj]{emulateapj}
\usepackage{color}
\usepackage[linktocpage]{hyperref}
\usepackage{graphicx,amssymb,amsmath,amsthm,amsfonts,epsfig,bm}
\usepackage{multirow}
{\renewcommand{\arraystretch}{1.5}
\begin{document}

\title{Gravitational-wave detection and parameter estimation for accreting black-hole binaries and their 
electromagnetic counterpart
}

\author{Andrea Caputo\altaffilmark{1}, Laura Sberna\altaffilmark{2}, Alexandre Toubiana\altaffilmark{3,4}, Stanislav 
Babak\altaffilmark{3,8},
Enrico Barausse\altaffilmark{5,6,4}, Sylvain Marsat\altaffilmark{3}, Paolo Pani\altaffilmark{7}}

\altaffiltext{1}{Institut de F\'{i}sica Corpuscular --
	CSIC/Universitat de Val\`{e}ncia, Parc Cient\'ific de Paterna.\\
	C/ Catedr\'atico Jos\'e Beltr\'an, 2 E-46980 Paterna (Valencia) - Spain}
\altaffiltext{2}{Perimeter Institute, 31 Caroline St N, Ontario, Canada}
\altaffiltext{3}{APC, AstroParticule et Cosmologie, Universit\'e Paris Diderot,
CNRS/IN2P3, CEA/Irfu, Observatoire de Paris, Sorbonne Paris Cit\'e,
10, rue Alice Domon et L\'eonie Duquet 75205 Paris Cedex 13, France}
\altaffiltext{4}{Institut d'Astrophysique de Paris, CNRS \& Sorbonne
 Universit\'es, UMR 7095, 98 bis bd Arago, 75014 Paris, France}
\altaffiltext{5}{SISSA, Via Bonomea 265, 34136 Trieste, Italy \& INFN, Sezione di Trieste}
\altaffiltext{6}{IFPU - Institute for Fundamental Physics of the Universe, Via Beirut 2, 34014 Trieste, Italy}
\altaffiltext{7}{Dipartimento di Fisica, ``Sapienza'' Universit\`a di Roma \& Sezione INFN Roma1, Piazzale Aldo Moro 5, 
00185, Roma, Italy.}
\altaffiltext{8}{Moscow Institute of Physics and Technology, Dolgoprudny, Moscow region, Russia}

\begin{abstract}
We study the impact of gas accretion on the orbital evolution of black-hole binaries initially at large separation in 
the band of the planned Laser Interferometer Space Antenna (LISA). We focus on two sources: (i)~stellar-origin 
black-hole binaries~(SOBHBs) that can migrate from the LISA band to the band of ground-based gravitational-wave~(GW) 
observatories within weeks/months; and (ii) intermediate-mass black-hole binaries~(IMBHBs) in the LISA band only.
Because of the large number of observable GW cycles,  
the phase evolution of these systems needs to be modeled to great accuracy to avoid
biasing the estimation of the source parameters.
Accretion affects the GW phase at negative 
($-4$) post-Newtonian order, being thus dominant for binaries at large separations.  
Accretion at the Eddington or at super-Eddington rate will leave a detectable imprint on the dynamics 
of SOBHBs. For super-Eddington rates and a 10-years mission, a multiwavelength strategy with LISA and a 
ground-based interferometer can detect about $10$ (a few) SOBHB events for which the accretion rate can be measured at $50\%$ ($10\%$) level. In all cases the sky position can be identified within much less than $0.4\,{\rm deg}^2$ uncertainty. 
Likewise, accretion at $\gtrsim 100\%$ of the Eddington rate can be measured in IMBHBs up to redshift 
$z\approx 0.1$, and the position of these sources can be identified within less than $0.01\,{\rm 
deg}^2$ uncertainty. 
Altogether, a detection of SOBHBs or IMBHBs would allow for targeted searches of electromagnetic 
counterparts to black-hole mergers in gas-rich environments with future X-ray detectors (such as Athena) and/or radio 
observatories (such as SKA). 
\end{abstract}


%%%%%%%%%%%%%%%%%%%%%%%%%%%%%%%%%%%%%%%%%%%%%%%%%%%%%%%%%%%%
\section{Introduction}
%%%%%%%%%%%%%%%%%%%%%%%%%%%%%%%%%%%%%%%%%%%%%%%%%%%%%%%%%%%%
Among the main gravitational-wave~(GW) sources detectable by the future Laser Interferometer Space Antenna 
(LISA)~\citep{Audley:2017drz} are binary black holes with relatively small masses, down to a few tens of solar masses~\citep{Sesana:2016ljz}.
LISA can detect these systems when they are still at large separations and thus probe their low-frequency dynamics. 
In more detail, these systems include: (i)~\emph{stellar-origin black hole binaries}~(SOBHBs) of a few tens up to $\sim 100 M_\odot$, whose
coalescences are also observed by terrestrial GW
detectors~\citep{LIGOScientific:2018jsj}; and, if they exist, (ii) \emph{intermediate mass black hole binaries}~(IMBHBs) with component 
masses in the range $(10^2,10^5)M_\odot$~\citep{Miller:2003sc}.

SOBHBs will be first observed in the LISA $\sim$ mHz band, and will then disappear for weeks/months 
before entering the $\gtrsim 1$ Hz band of ground detectors, where they merge~\citep{Sesana:2016ljz}.
Despite this frequency gap, piercing together the LISA low-frequency regime and the terrestrial high-frequency merger
will allow for effectively observing these systems for $10^5$--$10^6$ GW cycles. Therefore,
even small inaccuracies in modeling the GW phase evolution will bias the estimation of the parameters (and particularly
the merger time) or even prevent detection by LISA.

IMBHBs might be detected by LISA for the first time for a whole range of total masses and mass ratios, with the lighter 
binaries spending more time in band. 
While the existence of intermediate-mass
black holes has not been confirmed yet, several candidates exist~\citep[see e.g. ][for a review]{imbh},
and they might also provide seeds for the growth of the supermassive black holes
that are ubiquitously observed in the local universe~\citep[see e.g. ][]{imbh,latif}.
While their formation mechanism is unknown, proposed scenarios include direct collapse of massive first-generation, 
low-metallicity Population~III stars~\citep{Madau:2001sc,Schneider:2001bu,Ryu:2016dou,Kinugawa:2014zha}, runaway 
mergers of massive main sequence stars in
dense stellar 
clusters~\citep{Miller:2001ez,PortegiesZwart:2002iks,AtakanGurkan:2003hm,PortegiesZwart:2004ggg,Mapelli:2016vca}; 
accretion of residual gas onto stellar-origin black holes~\citep{Leigh:2013tqa}; and chemically homogeneous 
evolution~\citep{Marchant:2016wow}.

Both SOBHBs and IMBHBs offer the potential to constrain low-frequency
modifications of the phase evolution, {\it if} the latter are included in the GW templates used for the analysis in the 
LISA band.
Such low-frequency phase modifications may appear, e.g., if the dynamics of these systems
is governed by a theory extending/modifying general relativity~\citep{Barausse:2016eii,Carson:2019rda,Gnocchi:2019jzp}, 
or as a result of interactions (already within general relativity) of the binary with the surrounding gas, 
if the latter is present~\citep{Barausse:2014tra,Barausse:2014pra,Tamanini:2019usx,Cardoso:2019rou}.

There is currently no evidence that the SOBHBs
observed by GW detectors live in gas-rich environments -- and 
no electromagnetic~(EM) counterpart to these sources has been detected so far~\citep{Abbott:2016gcq}.
Binaries involving accreting stellar-origin black holes
are observed in X-rays~\citep{2003astro.ph..8020C}, but the accreting gas is provided by a stellar companion. 
However, gas may  be present earlier 
in the evolution of SOBHBs, and some of it may survive in the binary's surroundings. For instance, 
in the field-binary formation scenario~\citep{TheLIGOScientific:2016htt} for SOBHBs, gas plays a key role 
in the common envelope phase, although the latter typically precedes the merger  by several Myr.
Also note that SOBHBs may form preferentially in the gas-rich nuclear regions surrounding 
AGNs~\citep{2018ApJ...866...66M} -- as a result e.g. of Kozai-Lidov resonances~\citep{2012ApJ...757...27A} or simply 
fragmentation/instabilities of the AGN
accretion disk~\citep{2017MNRAS.464..946S}. 
Furthermore, accretion onto stellar-origin or intermediate-mass black holes has been proposed as an explanation for
ultra-luminous X-ray sources~\citep[see e.g. ][]{2004ApJ...614L.117M}. Accretion, in combination with
mergers, is also thought to be the main channel via which 
black hole seeds evolve into the supermassive black holes we observe today.

Therefore, at least some SOBHBs or IMBHBs may still
be accreting matter in the LISA band and perhaps even at merger. The accretion-driven EM
emission  may not have been detected  because these sources are too far\footnote{Note for instance
that accreting black holes in X-ray binaries are mostly observed in the Galaxy, with only a few observed in
nearby galaxies. Among the latter, the farthest is M83~\citep{Ducci:2013xwa} 
which is only $\sim 4.5$ Mpc away, 
vs the several hundred
Mpc of the LIGO/Virgo SOBHBs~\citep{LIGOScientific:2018jsj}.}, because accretion
is radiatively inefficient~\citep{2002apa..book.....F}, or because the  sky position uncertainty provided by GWs is too
large for follow-up campaigns. 
Note also that LISA is expected to detect up to several tens of SOBHBs~\citep{Sesana:2016ljz,Tamanini:2019usx}. 
If only one such system
were accreting, and if the {\it possibility} for accretion were not included
in the GW templates used for the analysis, the parameter estimation may mistakenly point towards a
modification of general relativity~\citep{Barausse:2016eii,Carson:2019rda,Gnocchi:2019jzp}
--~a claim that would have groundbreaking effects on physics. 
Furthermore, LISA may provide an accurate sky localization for these sources, thus increasing the chances of detecting 
a putative EM counterpart, with important implications for multimessenger astronomy and cosmology.

With these motivations, in this work we analyze the effect of gas accretion
on standalone IMBHB LISA detections and on joint LISA+ground multiwavelength SOBHB observations.
We find that accretion introduces a $-4$ Post-Newtonian (PN) correction to the phase\footnote{In the GW phase, a $n$PN correction scales as $(v/c)^{2n}\sim f^{2n/3}$ ($v$ and $f$ being the binary's 
orbital velocity and the GW frequency) 
relative to the leading-order general relativistic term.}, thus potentially dominating over the
GW-driven evolution at low frequencies~\citep[see also e.g.][]{Holgado:2019ndl}. 
The systems we consider will be driven by gravitational wave emission, with accretion acting as a perturbative correction and therefore leaving an imprint on the GW phasing. 
We explore the consequences of this
fact for GW parameter estimation, i.e. we assess both with what uncertainty
the accretion rate can be recovered when the possibility for accretion is included in the templates, and
how much the estimate of the binary parameters will be biased if it is not. 
We also look at the prospects of identifying the EM emission from
accreting SOBHBs and IMBHBs with observational facilities  available when LISA flies.

In Section~\ref{sec:shift} we begin by summarizing the effect of accretion on the GW waveform and on the binary evolution. In Section~\ref{sec:meas} we describe how we generate astrophysical catalogues and simulate future detections. We present our results in Section~\ref{sec:res} and we summarize them in Section~\ref{sec:disc}.
We use geometrized units in which $G=c=1$. We denote the total mass by $ M = m_1+ m_2 $, the reduced mass by $ \mu=m_1 m_2 / M $, and the chirp mass by $ \mathcal{M} =  \left( \mu^3 M^2\right)^{1/5}  $.

%%%%%%%%%%%%%%%%%%%%%%%%%%%%%%%%%%%%%%%%%%%%%%%%%%%%%%%%%%%%
\section{Shift of the merger time and waveform corrections due to accretion}\label{sec:shift}
%%%%%%%%%%%%%%%%%%%%%%%%%%%%%%%%%%%%%%%%%%%%%%%%%%%%%%%%%%%%
Let us parametrize
the mass accretion rate of each component of a (circular) black-hole binary (with masses $m_i$, $i=1,2$) by the Eddington 
ratio
\begin{equation}
	f_{{\rm Edd},i}=\frac{\dot{m}_i}{\dot{m}_{\rm Edd}}\,,
	\label{eddmass}
\end{equation} 
where $\dot{m}_{\rm Edd} \simeq 2.2 \times 10^{-8} \left( \frac{m_i}{M_{\odot }} \right) M_{\odot} \, {\rm 
yr}^{-1}$ is the Eddington accretion rate 
(obtained from the Eddington luminosity assuming radiative efficiency $\eta=0.1$). 
Since the accretion timescale exceeds the dynamical timescales of the binary when the latter is in the 
frequency band of LISA or ground detectors, the effect on the phase can be computed using the stationary phase 
approximation, and to the leading order at low frequencies
% \footnote{\textbf{Note that Eq.~\eqref{phase} gives only the 
% leading order term of the phase correction due to accretion. In our computations we took into account also 
% the subleading terms that become relevant at higher frequencies, see Appendix~\ref{app:waveform} for the full 
% expression.}}
it reads (c.f. derivation in Appendix~\ref{app:waveform})
\begin{equation}\label{phase}
\phi_{\rm acc} \sim \alpha f_{\rm Edd}\,f^{-13/3} \,,
\end{equation} 
%%%
where $f$ is the GW frequency and $\alpha$ is a coefficient that depends on the binary parameters.
Since the leading-order term in the phase in vacuum is $\sim 
f^{-5/3}$~\citep{Maggiore}, this is a -4PN term, which dominates the binary evolution at low frequencies.
In the frequency range of LISA observations, due to the smallness of the prefactor, this term will be a small correction 
to the vacuum GW phase. In other words, our SOBHB and IMBHB sources will emit GWs well above the frequency at which 
accretion becomes subdominant,
\begin{equation}\label{key}
f\gg 1.1 \times 10^{-4} \, \left(\frac{f_{\rm Edd}}{1}\right)^{3/8} \left(\frac{\mathcal{M}}{10 \,  M_{\odot}}\right)^{-5/8}\, {\rm Hz} \, ,
\end{equation} 
see  Appendix~\ref{app:waveform}.

As a result of accretion, the phase evolution accelerates and the binary merges earlier (i.e. in less time and 
in fewer GW cycles) than in vacuum. Note that in this work we neglect for simplicity the hydrodynamic drag
produced by the transfer of linear momentum by the accreting gas~\citep{Barausse:2007dy, Chen:2019jde}, which would further contribute to the shift of the merger time (see Appendix~\ref{app:waveform}).

In Fig.~\ref{fig:dphiacc} we show the  time $T$ needed for a SOBHB to enter the band of ground detectors (top panel), the 
time difference $ \Delta T$ in the merger time induced by accretion (middle panel), and the 
difference $\Delta\phi$ in the  total (accumulated) GW phase  
 due to accretion (bottom panel), as functions 
of the initial GW frequency in the LISA band and for
various SOBHB masses. All these quantities can be computed either numerically solving 
Eq.~\eqref{eq:equationaccr} or using the perturbative expansions in Eqs.~(\ref{t1acc}--\ref{phaseA}). The two approaches 
are in excellent agreement because the contribution of accretion is subdominant in all cases.

As a useful rule of thumb, time differences $\Delta T>10\,{\rm s}$~\citep{Sesana:2016ljz}
and phase differences $\gtrsim 1$ rad~\citep{Flanagan:1997kp,Lindblom:2008cm} are large enough to be detectable. 

For low initial frequency, the 
effect of accretion on $\Delta T$ and on the phase is stronger, but the time $T$  is also very large, i.e. multi-wavelength
observations will be impossible in practice. One may try to detect accretion with LISA data alone, but note
that the mission's duration will not exceed 10 yr (with a nominal duration of 4 yr), due to
the finite consumables carried by the spacecraft. 
For these reasons, we mark in Fig.~\ref{fig:dphiacc} the phase and time differences for a SOBHB 
that enters the band of ground detectors in 10 (4) yr by full (empty) circles. The part of the curves to the right of these circles 
then corresponds to  $T<10\,{\rm yr}$ 
($T<4\,{\rm yr}$), which would make a joint LISA+ground detection possible in pratical terms.
Overall, the results of Fig.~\ref{fig:dphiacc} (which scale linearly with $f_{{\rm Edd}}$) suggest that only $f_{\rm Edd}>0.1$ would give a potentially detectable 
effect, i.e. $\Delta T>10\,{\rm s}$ and $\Delta\phi\gtrsim 1$. We will verify this with more rigorous techniques in 
the following.
\begin{figure}[ht!]
	\centering
	        \includegraphics[width=0.46\textwidth]{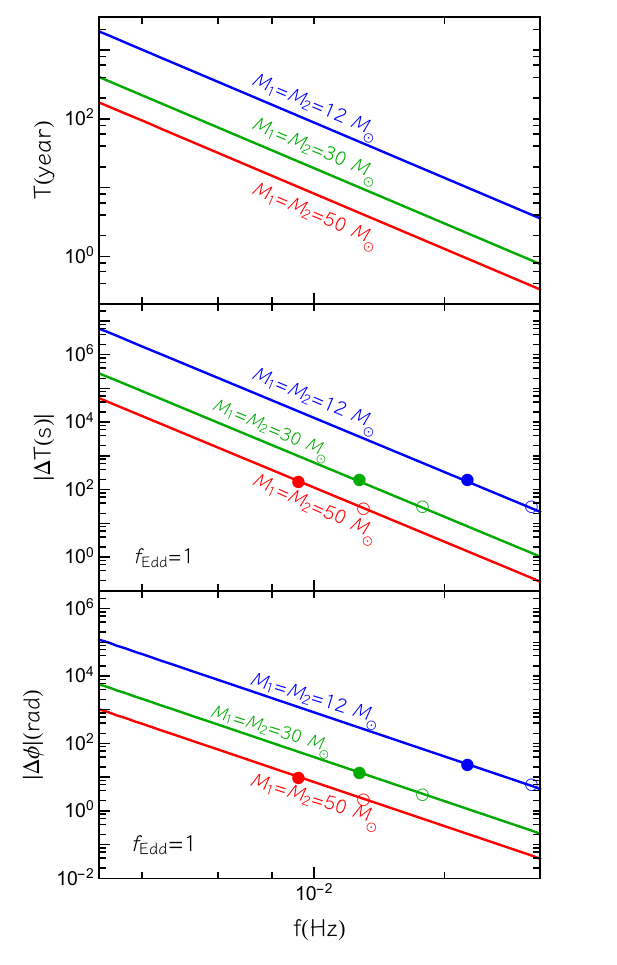}\\
	\caption{Time needed to enter the band of terrestrial detectors neglecting accretion~(top), time difference caused by 
accretion~(middle), and corresponding GW phase difference (bottom), as functions of the initial GW frequency for three 
equal-mass SOBHBs. 
We choose $f_{\rm Edd}=1$ as a reference, since the 
 time  and phase differences scale linearly with $f_{\rm Edd}$. The full (empty) circles mark the points corresponding to 
$T=10\,{\rm 
yr}$ ($T=4\,{\rm yr}$). The systems to the right of the full (empty) circles therefore have $T<10\,{\rm yr}$ 
($T<4\,{\rm yr}$). 
}\label{fig:dphiacc}
\end{figure}

%%%%%%%%%%%%%%%%%%%%%%%%%%%%%%%%%%%%%%%%%%%%%%%%%%%%%%%%%%%%%%%%%%%%%%%%%
\section{Measuring accretion effects for SOBHBs and IMBHBs}\label{sec:meas}
%%%%%%%%%%%%%%%%%%%%%%%%%%%%%%%%%%%%%%%%%%%%%%%%%%%%%%%%%%%%%%%%%%%%%%%%%

In order to quantify the ability of multiband SOBHB detections and standalone IMBHB observations to constrain the 
accretion model, we perform two analyses: (i)~A simple Fisher matrix analysis to explore the whole parameter space, 
and (ii)~a more refined Monte Carlo 
Markov Chain~(MCMC) analysis for the best candidate events. Note that the Fisher matrix analysis 
is only valid for large 
signal-to-noise ratios~(SNRs)~\citep{Vallisneri:2007ev}. Therefore, we expect it to provide
only qualitatively correct results for SOBHBs in the LISA 
band (for which the SNR is at most $15-20$ in the most optimistic cases, see 
below). Nevertheless, we expect the Fisher matrix analysis to be accurate for the IMBHBs we consider, for which ${\rm SNR}={\cal O}(100)$.

In both the Fisher and MCMC analyses we only account for the contribution 
due to accretion in the GW phase, and neglect the subleading contribution to the amplitude. 
Since accretion is important at low frequency, high-order PN terms (including the spin) should be irrelevant for 
our analysis, but we include them for completeness and to estimate possible correlations.

For simplicity, in the Fisher 
analysis we also neglect the motion of the antenna during the observation. This is instead included in the MCMC analysis, in 
order to estimate the ability to localize the source in the sky and measure the accretion rate at the same time.

Finally, we consider two situations: one (referred to as \emph{LISA+Earth}) in which we simulate a multiband SOBHB detection 
(LISA combined with a ground-based interferometer) and another (referred to as \emph{LISA-only}) in which we simulate a 
standalone (either SOBHB or IMBHB) detection by LISA. In the \emph{LISA+Earth} case, to simulate a multiband detection 
one can follow two options: combine statistically the noise curves of LISA with that of a given ground-based detector or, 
alternatively (but less rigorously), assume that the merger time can be computed independently by the ground-based detector, so that the 
dimension of the parameter space of the analysis is effectively reduced.
 In the Fisher analysis, we follow the latter, simpler approach,  and we therefore effectively
remove the merger time from the template parameters in the \emph{LISA+Earth} case. In the MCMC analysis we keep $t_c$ as a free parameter, restricting it by using a narrow prior.
In all cases we adopt the LISA noise curve reported by~\citet{Audley:2017drz}, 
whose high frequency part is based on a single link optical measurement
system noise of 10 pm/$\sqrt{\mbox{Hz}}$.

%%%%%%%%%
\subsection{Fisher analysis and event rates for SOBHBs}
%%%%%%%%%
In the Fisher analysis we adopt a TaylorF2 template approximant for spinning binaries up to $3.5$PN order~\citep{Droz_1999}, with the addition of the leading-order accretion term presented in Eq.~\eqref{phase}. 
Therefore, our GW template for the Fisher analysis has \emph{seven} parameters (masses, merger time and phase, the two 
dimensionless spins $\chi_{1,2}$, besides the 
Eddington accretion ratio $f_{\rm Edd}$). 

Given a waveform template $h(\vec{\zeta},f)$ in the frequency domain and a set of waveform parameters $\vec{\zeta}$, 
the error associated with the measurement of parameter $\zeta^a$ (with all other parameters marginalized upon) 
is $\sigma_{a}=\sqrt{\Sigma^{aa}}$, where the covariance matrix $\Sigma^{ab}$ is given by the 
inverse of the Fisher matrix, $\Gamma_{ab}=\left(\partial_{\zeta^a} h \vert \partial_{\zeta^b} 
h\right)_{\vec{\zeta}=\vec{\zeta}_0}$. Here, $\vec{\zeta}_0$ are the injected values of the parameters, and the 
inner product is defined by
\begin{equation}\label{inner}
(g\vert h)=4\,{\rm Re}\int_{f_{\rm 0}}^{f_{\rm max}}df\,\frac{\tilde h(f)\tilde g^{\star}(f)}{S_{h}(f)}\ ,
\end{equation}
where $S_h(f)$ is the detector noise spectral density.

While the number (and the very existence) of IMBHBs in the LISA band is very uncertain, our  Fisher-matrix analysis,
coupled with simulated astrophysical populations calibrated to the LIGO/Virgo data, can easily provide 
estimates of the number of SOBHBs detectable by LISA for which accretion can be measured. 
The intrinsic number of SOBHBs merging per (detector-frame) unit time and (source-frame) masses is given by~\citep{Hartwig:2016nde}
\begin{equation}\label{rates}
\frac{\mathrm{d} \dot{N}}{\mathrm{d} m_1 \mathrm{d} m_2}=\int \mathrm{d} z \,
R \frac{\mathrm{d}t}{\mathrm{d} z} 
\frac{\mathrm{d}^2 p}{\mathrm{d} m_1 \mathrm{d} m_2} 4 \pi d_C^2,
\end{equation} 
where $ d_{C}  $ is the comoving distance, $ R = 53.2 \, \text{Gpc}^{-3} \text{yr}^{-1}$
is the best estimate for the intrisic merger rate measured by the first and second LIGO/Virgo runs~\citep{LIGOScientific:2018jsj},
the probability distribution function for the source-frame masses -- ${\mathrm{d}^2 p}/{\mathrm{d} m_1 \mathrm{d} m_2}$ --
is given by ``model B'' of \citet{LIGOScientific:2018jsj}, while $$\frac{\mathrm{d}t}{\mathrm{d} z}=\frac{1}{H_0(1+z)\sqrt{\Omega_m(1 + z)^3 +  \Omega_{\Lambda}}}$$ 
is computed using our fiducial cosmology $H_0 = 67.9 \,\text{km}/\text{s}/\text{Mpc}  $, 
$ \Omega_m =0.306 $, $ \Omega_{\Lambda} = 0.694$~\citep{Ade:2015xua}.
In order to obtain synthetic astrophysical catalogues of merging as well as inspiraling sources, we use Eq.~\eqref{rates} to
simulate mergers in a period much longer than the LISA mission duration, by
assuming a uniformly distributed merger time $t_c$. The latter can be easily
converted into the initial GW frequency $f_0=[5/(256 \, t_c)]^{3/8} \mathcal{M}^{-5/8}/\pi$, where $f_0$, $t_c$, and the 
chirp mass $\mathcal{M}$ must be computed in the same (detector- or source-) frame.

We constrain the comoving distance in the range $d_C \le 2 \, {\rm Gpc}$ and the initial source frame GW frequency in the range 
$f_0\in[4 \,{\rm mHz} ,10 \,{\rm Hz}]$. For the chosen mass model we generate $20$~realizations, and for each 
realization we consider two LISA mission durations ($ 4 $ or $ 10 $~yr), for a total of $40$~catalogues.

In the \emph{LISA-only} case for SOBHBs, we assume that a single event within the catalogue is detected if either of the following conditions occurs~\citep{Moore:2019pke,Tamanini:2019usx}
\begin{align}%\label{rules2}
t_c< 100\,{\rm yr}  \text{ and SNR}&\ge 15,\text{ or} \nonumber \\
t_c> 100\,{\rm yr} \text{ and SNR}&\ge 10\,	, \nonumber	
\end{align} 

where the latter SNR threshold is lower because binaries with long merger times are accurately described by Newtonian 
waveforms in the LISA band~\citep{Mangiagli:2018kpu} and can be therefore detected by a different search strategy~\citep{Tamanini:2019usx}, akin e.g. to
the one used for white-dwarf binaries.

In the \emph{LISA+Earth} case, the SNR threshold is lower for events that can be detected on Earth~\citep{Moore:2019pke,Tamanini:2019usx}
\begin{align}\label{rules}
t_c< 10\,{\rm yr} \text{ and SNR}&\ge 9.5\,.	
\end{align} 
These events would indeed be detected through an archival search following their ground-based detection.
%%%

%%%%%%%%%%%%%%%%%%%%%%%%%%%%%%%%%%%%%%%%%%%%%%%%%%%%%%%%%%%%%%%%%%%%
\subsection{MCMC with sky localization and antenna motion}\label{mcmc}
%%%%%%%%%%%%%%%%%%%%%%%%%%%%%%%%%%%%%%%%%%%%%%%%%%%%%%%%%%%%%%%%%%%%

For the MCMC analysis we adopt the PhenomD template~\citep{Husa:2015iqa,KHH16} with the inclusion of the phase term due to accretion. 
In this case we also account for the motion of the antenna during the observation. More specifically, the standard part of the GW template is the same as in~\citet{Tamanini:2019usx} and contains \emph{five} additional 
parameters besides those adopted for the Fisher matrix analysis: two angles identifying the 
source position with respect to the detector ($\phi$,$\theta$), the GW polarization ($\psi$), the inclination of the system ($\iota$), and the luminosity distance ($d_L$).

In the \emph{LISA-only} scenario we use $f_0$ as sampling parameter and assume a flat prior for it. In the \emph{LISA+Earth} scenario we use $t_c$ with a Gaussian prior centered around the true value with width $\sigma_{t_c}=10^{-3} \, {\rm s}$ which models the fact that $t_c$ can be measured with great precision in this scenario. For IMBHBs, we consider a single \emph{LISA-only} scenario.

{\renewcommand{\arraystretch}{1.3}
\begin{table*}[t]
\caption{Number of detectable SOBHB events for various configurations. ``All'' stands for the total number of 
detectable events, whereas $100\%$, $50\%$, and $10\%$ stand for the 
number of events for which $f_{\rm Edd}$ is measured with a relative error of $100\%$, $50\%$, and $10\%$, 
respectively, according to the Fisher analysis. All numbers are averaged over $20$ catalogues and presented with $1\, 
\sigma$ errors. Super-Eddington accretion will be detectable for a good fraction of multiband events if the LISA mission duration is 10 years.}
\label{tab:events}
\centering
\begin{tabular}{|*{11}{c|}}
	\cline{2-11}
	\multicolumn{1}{c|}{}          & \multicolumn{5}{|c|}{ LISA+Earth} & \multicolumn{5}{c|}{LISA-only} \\ \hline
	\multicolumn{1}{|c|}{Duration} & All  & $f_{\rm Edd}$    & 100\%   & 50\%  & 10\%     & All   & $f_{\rm Edd}$   & 100\%   & 50\%  & 10\%   \\  \hline
	\multicolumn{1}{|c|}{\multirow{2}{*}{4 yr }} &  \multirow{2}{*}{88 $\pm$ 8 }  %&  0.1      &  0 & 0  &    0     
	&  1      &  0.1 $\pm$  0.2  & 0  &    0   
	&  \multirow{2}{*}{77    $\pm$ 8} %& 0.1   &  0  & 0 & 0   \\ \cline{3-6} \cline{8-11}
	 & 1 &  0  & 0  &    0    \\ \cline{3-6} \cline{8-11}
	& &  10     &    4.1 $ \pm $ 2.3 & 1.7 $ \pm $ 1.2  &  0.1 $ \pm $ 0.2       
	&     & 10 &  1.6 $ \pm $ 1.4  &  0.6 $ \pm $ 0.6 &  0      \\
	%%%%%%
	\hline\hline  \multicolumn{1}{|c|}{ \multirow{2}{*}{10 yr} }    & \multirow{2}{*}{207  $\pm$ 11} & 
	 1 &  5.2  $\pm$ 1.9  &   1.1 $\pm$ 1.2  &  0.1 $\pm$ 0.2     
	&  \multirow{2}{*}{182    $\pm$ 10} &     
	 1 & 1.5 $ \pm $ 1.2  &   0.4 $ \pm $ 0.7  &  0
	\\ \cline{3-6} \cline{8-11}
	& &  10     &      36 $\pm$ 4   & 32 $\pm$  3 &    5.2 $\pm$ 1.9      
	&     & 10 &  11 $\pm$ 3  & 9.5 $ \pm $ 2.7  &    1.5 $ \pm $ 1.2    \\  \cline{1-11}
\end{tabular}
\end{table*}

When including the source location, different realizations of the angles for the same astrophysical system yield 
different SNRs. This affects the precision within which one can recover the parameters of the source, including the sky position itself and $f_{\rm Edd}$. In order to cross-check results obtained with our Fisher matrix analysis and to quantify this variability, we select from the catalogue an astrophysical system for which the accretion parameter can be measured precisely through the Fisher matrix approach, and draw 
three different realizations of ($\phi$, $\theta$, $\psi$, $\iota$) yielding a low SNR $\sim9$, a medium SNR 
$\sim15$, and a high SNR $\sim20$, respectively. The medium SNR system is chosen so that its SNR is close to 
the value obtained by averaging over the angles.

For IMBHBs, we consider two systems (see details in the next section): one merging in the LIGO/Virgo band, and one with higher masses, merging at lower frequencies. We choose the inital frequency so that both systems merge in $10\,{\rm yr}$, the longest possible LISA mission duration. 

For each of these systems we perform a full Bayesian analysis (see Appendix~\ref{app:MCMC}).
We simulate GW data $d(f)$ as it would be measured by LISA, 
computing the response of the detector (accounting for the constellation's motion) by following~\citet{Marsat:2018oam}. We work in the zero noise approximation
in order to speed up the computation. Adding noise to the GW signal should not affect the parameter estimation 
drastically, leading mostly to a displacement in the maximum of the parameter distribution~\citep{Rodriguez:2013oaa}.

We perform two different analyses: in the first one we generate data with a non-zero value for $f_{\rm 
Edd}$ and include it as a free parameter in the Bayesian analysis, in order to estimate with what precision 
it can be recovered. In the second case, data are also generated with a non-zero value of $f_{\rm 
Edd}$, but when doing the analysis we set $f_{\rm Edd}=0$ in the templates, in order to measure the bias in the 
parameter estimation. 
In all cases, the posterior distribution is computed using Bayes' theorem. Additional details are given in Appendix~\ref{app:MCMC}.

%%%%%%%%%%%%%%%%%%%%%%%%%%%%%%%%%%%%%%%%%%%%%%%%%%%%%%%%%%%%%%%%%%%%%%%%%%%%%%%%%%%%%%%%%%
\section{Results}\label{sec:res}
%%%%%%%%%%%%%%%%%%%%%%%%%%%%%%%%%%%%%%%%%%%%%%%%%%%%%%%%%%%%%%%%%%%%%%%%%%%%%%%%%%%%%%%%%%
%%%%%%%%%%%%%%%%%%%%%%%%%%%%%%%%%%%%%%%%%%%%%
\subsection{Event rates for SOBHBs}

%%%%%%%%%%%%%%%%%%%%%%%%%%%%%%%%%%%%%%%%%%%%%
%
For the simulated astrophysical populations we first use a Fisher matrix analysis to quantify the possibility 
to measure $f_{\rm Edd}$ at a given precision. 
Table~\ref{tab:events} shows the average number of detected SOBHBs, and the number of SOBHBs for which 
$f_{\rm Edd}$ can be measured within a given precision. The results are obtained by 
averaging the Fisher matrix over sky position and source inclination (while neglecting,
as already mentioned, the LISA constellation's motion), for different injected values of $f_{\rm Edd}$.
Our results for the total number of detected events are consistent with~\citet{Tamanini:2019usx, Sesana:2016ljz}. 

In particular, for the \emph{LISA+Earth} case and a $10\,{\rm yr}$ mission, super-Eddington accretion $ f_{\rm Edd}\approx 10 $ can be measured 
within $50\%$ precision in about $15\%$ of the total detectable events ($\approx 200$), while a measurement within $10\%$ 
is only possible in $\sim2\% $ of the events. Note that the statistical errors scale approximately linearly with $f_{\rm Edd}$. Therefore, when injecting a 
lower accretion rate the number of events for which accretion is measurable is significantly 
smaller. For example, $f_{\rm Edd}=1$ is marginally detectable in $\lesssim 1$ event in the most optimistic scenario, whereas smaller values of the accretion rates are not measurable.

As expected, a multiband observation improves the measurements of a negative-PN term, including the -4PN term due  to accretion: the event rates for the \emph{LISA-only} case are thus smaller by a factor of a few relative to the \emph{LISA+Earth} case.

%%%%%%%%%%%%%%%%%%%%%%%%%%%%%%%%%%%%%%%%%%%%%
\subsection{Measuring accretion and sky localization}
%%%%%%%%%%%%%%%%%%%%%%%%%%%%%%%%%%%%%%%%%%%%%

% \subsubsection{Detection}
For our MCMC analysis we select one representative SOBHB system from our synthetic astrophysical catalogues, and choose two 
optimistic IMBHB systems on the basis of a Fisher matrix analysis spannning the parameter space, i.e. 
the errors on $f_{\rm Edd}$ provided by the chosen IMBHBs are roughly the smallest throughout the parameter space. In more detail, the systems
that we consider are 
\begin{itemize}
 \item A SOBHB with $m_1= 42.1 \  M_{\odot}$, $m_2=39.8 \ 
M_{\odot}$, $\chi_1=0.008$, $\chi_2=0.44$, at a distance $ d_L =416 \ {\rm Mpc} $;
 \item An IMBHB with $m_1=315 \  M_{\odot}$, $m_2=284 \ 
M_{\odot}$, $\chi_1=0.9$, $\chi_2=0.85$, referred to as ``light IMBHB'';
 \item Another IMBHB with $m_1=1000\  M_{\odot}$, $m_2=900 \ 
M_{\odot}$, $\chi_1=0.9$, $\chi_2=0.85$, referred to as ``heavy IMBHB''. 
\end{itemize}
%%%
For all three sources we set $t_c\approx 10$ yr.
We study the IMBHB systems at two different redshifts, $z=0.1$ and $z=0.5$, in order to estimate up to what distance  the presence of accretion in the binary would be detectable. The IMBHBs' masses are in the source frame and are kept fixed when the redshift is changed. 

For each realization of the angles $(\theta,\phi,\iota,\psi)$, we compute and sample the posterior distribution as explained in 
Appendix~\ref{app:MCMC}. 
As expected, the precision of the parameter measurements increases with the SNR. We find that the 
accretion parameter is strongly correlated with the intrinsic parameters of the source ($\mathcal{M}$, $\mu/M$, 
$f_0$, $\chi_s$, $\chi_a$), where $\chi_s$ and $\chi_a$ are defined in Appendix~\ref{app:MCMC}.

\begin{figure*}[ht!]
	\centering
	        \includegraphics[width=0.8\textwidth]{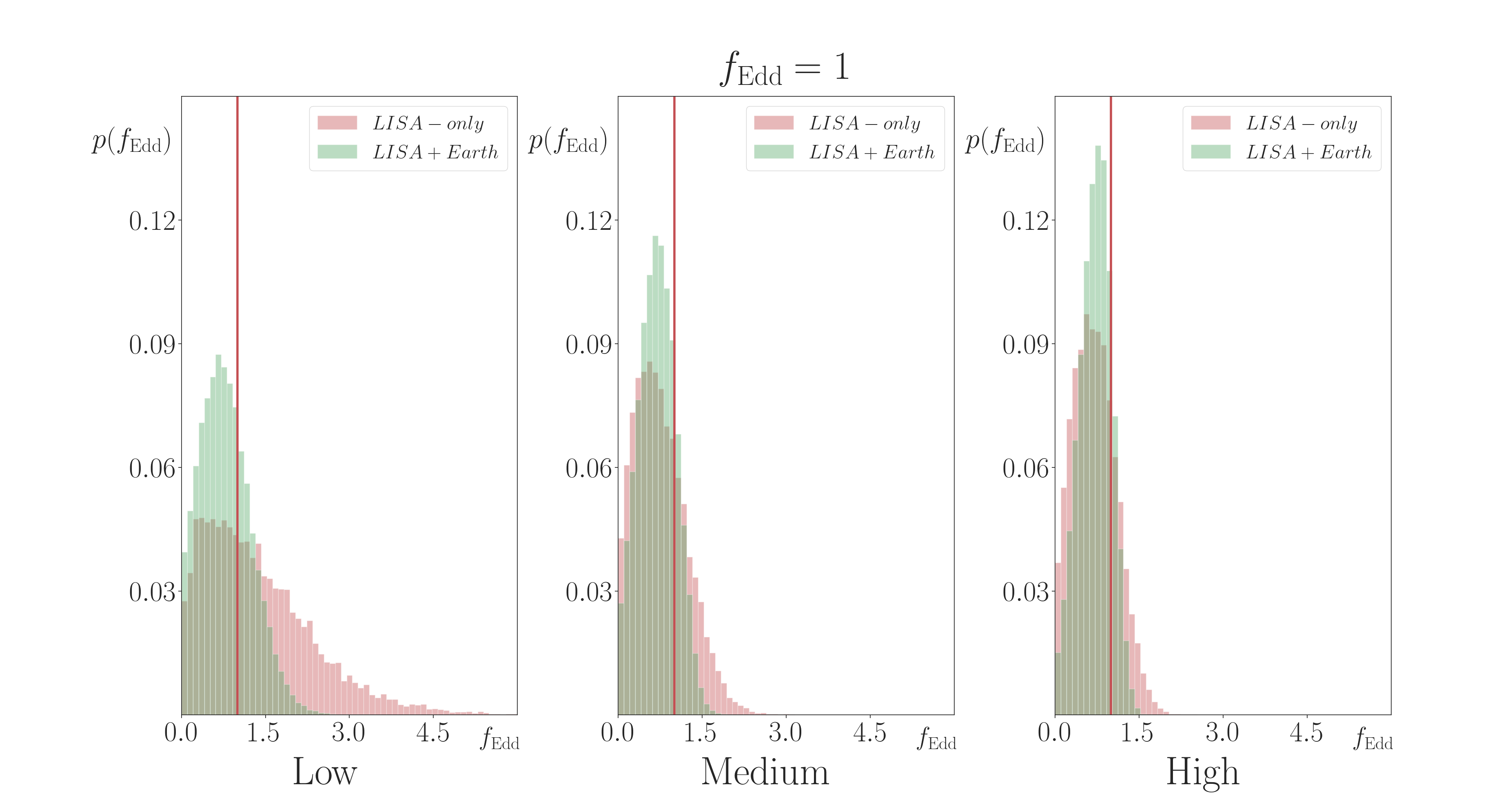}
	\caption{Marginalized distributions of $f_{\rm Edd}$ for the considered SOBHB for
various realizations of the angles, and in the scenarios \emph{LISA+Earth} (green) and \emph{LISA-only} (red). Red lines 
indicate the injected value of the accretion rate, $f_{\rm Edd}=1$. In the \emph{LISA+Earth} scenario and for higher SNR the marginalized distribution is strongly peaked but still consistent with $ f_{Edd}=0 $. }\label{fig:distr1}
\end{figure*}

In Fig.~\ref{fig:distr1} we show the marginalized distributions of $f_{\rm Edd}$ for the chosen SOBHB system, 
for various SNRs and for an injected value of $f_{\rm Edd}=1$.

For the high and the medium SNR cases, already in the \emph{LISA-only} scenario the posteriors indicate the presence of accretion. 
The marginalized distribution for $ f_{\rm Edd} $ can be compared with those obtained when constraining modifications of GR (some of which affect the vacuum waveform in a similar fashion as accretion, i.e. at negative PN orders) in the parameterized post Einsteinian framework~\citep{Yunes:2009ke}. 
In that case, as discussed in an  upcoming paper \citep{Toubiana20_2}, the marginalized distribution of the non GR-parameters is mostly flat 
up to a threshold (representing the upper bound that can be placed on the parameters under scrutiny), and then goes to $0$. In contrast, we see in Fig.~\ref{fig:distr1} that for high and medium SNR in the \emph{LISA-only} scenario, the distribution peaks at some nonvanishing value, indicating the presence of a non-zero modification to the vacuum waveform.

\begin{figure*}[ht!]
	\centering
	        \includegraphics[width=0.8\textwidth]{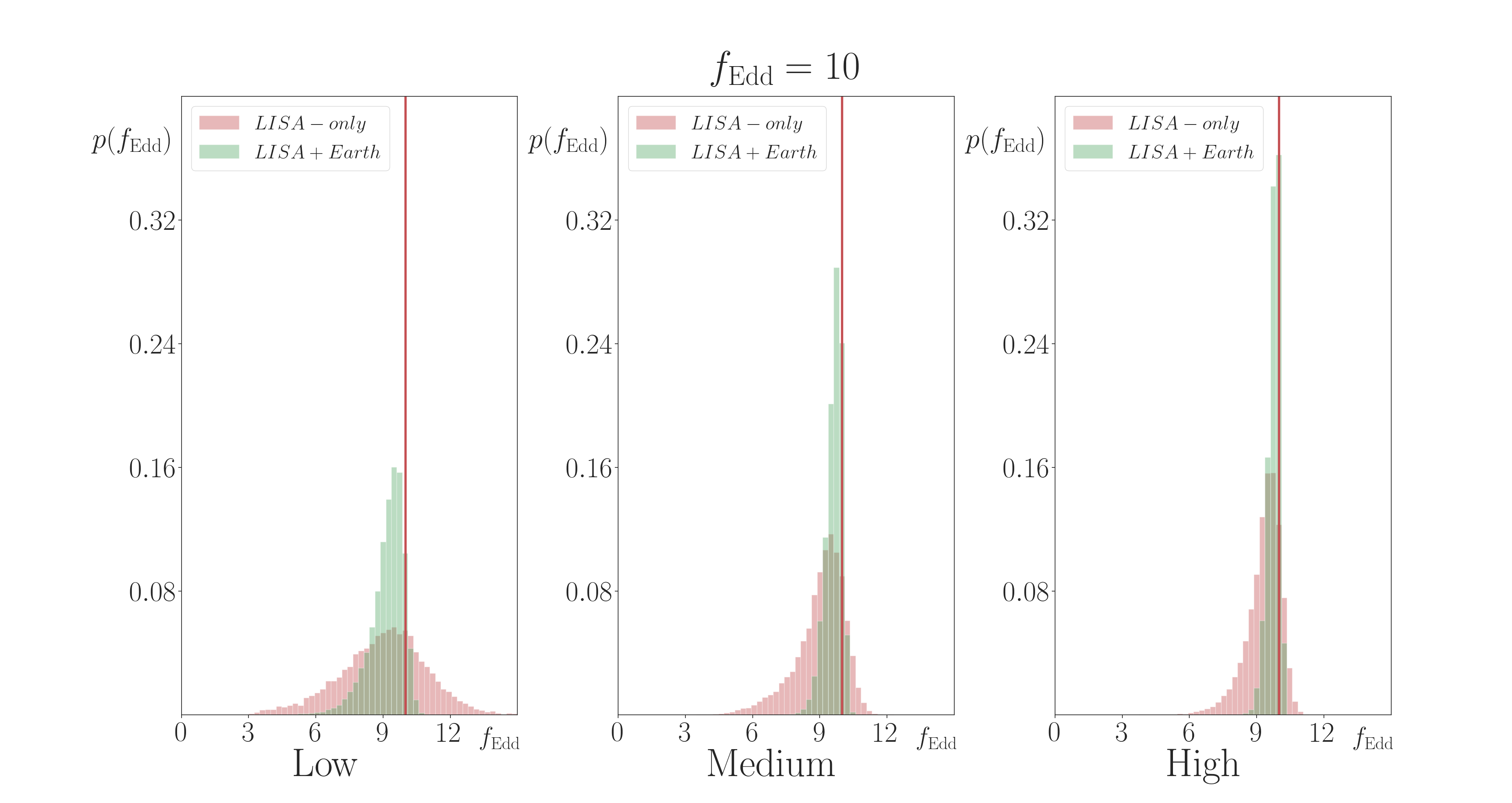}\
	\caption{
	Same as Fig.~\ref{fig:distr1} but for an injected value of $f_{\rm Edd}=10$. Accretion is detected in all scenarios and for any angle realization.
}
\label{fig:distr10}
\end{figure*}

In Fig.~\ref{fig:distr10} we show the same as in Fig.~\ref{fig:distr1}, but for an injected value of $f_{\rm Edd}=10$. 
This high accretion rate can be detected more easily even in the low SNR case and in the \emph{LISA-only} scenario, since in this case $f_{\rm Edd}=0$ is outside the support of the distribution.
Thus, for super-Eddington accreting binaries in the LISA band, there is a concrete chance to detect the effect of high rates of accretion on the waveform for most SOBHB events.

In Table~\ref{res_det} we show the recovered $68 \%$ confidence intervals~(CI) and median values for 
$f_{\rm Edd}$ and the sky localization ($\Delta \Omega$). 
In the $f_{\rm Edd}=1$ case, since the distribution is leaning against the boundary of the 
prior (see Fig.~\ref{fig:distr1}) we define the $68\%$ CI for $f_{\rm Edd}$ by taking the lower $68\%$ values. 
Instead, in the $f_{\rm Edd}=10$ case, the interval is centered around the median values. 
The marginalized distributions for $\Delta \Omega$ are approximately Gaussian and are centered around the injected value. 
Thus, we define the solid angle as~\citep{Cutler:1997ta}:
\begin{equation}
 \Delta \Omega=2\pi \sqrt{(\Sigma^{\phi,\phi}) (\Sigma^{\cos(\theta),\cos(\theta)})-(\Sigma^{\phi,\cos(\theta)})^2}\,.
\end{equation}

We show the same quantities for our IMBHB events in Table~\ref{res_det_imbh}. There, in the case $f_{\rm Edd}=1$, we define the $68 \%$ CI for $f_{\rm Edd}$ centered on the median, and in the case $f_{\rm Edd}=0.1$ we define it by taking the lower $68\%$ values.
In all cases considered here, the error on the sky 
localization is much smaller than the nomimal field of 
view of future X-ray and radio missions, potentially allowing for the detection of electromagnetic counterparts. We will discuss this possibility in Sec.~\ref{sec:multi}.

{
\renewcommand{\arraystretch}{1.5}
{\centering
  \begin{table}[th!] 
   \begin{tabular}{c|c|c|c|c|}
   \cline{2-5}
   
   & \multicolumn{2}{|c|}{$f_{\rm Edd}^{\rm injected}=1$}  & \multicolumn{2}{|c|}{$f_{\rm Edd}^{\rm injected}=10$}    
\\
   
   \cline{2-5}
   
   & $f_{\rm Edd}$ & $\Delta \Omega$ (${\rm deg}^2$) & $f_{\rm Edd}$ & $\Delta \Omega$ (${\rm deg}^2$)\\
   
   \hline
  
  \multicolumn{1}{|c|}{High SNR}  & $0.68_{-0.68}^{+0.02}$  & $0.14$ &$9.46_{-0.83}^{+0.53}$ & $0.14$  \\

  \hline

 \multicolumn{1}{|c|}{Medium SNR} & $0.70_{-0.70}^{+0.25}$ & $0.06$ & $9.23_{-1.23}^{+0.77}$ & $0.06$ \\

\hline
  
  \multicolumn{1}{|c|}{Low SNR} & $1.18_{-1.18}^{+0.50}$ & $0.33$ & $8.82_{-2.96}^{+1.85}$ & $0.34$ \\
  
  \hline

   \multicolumn{1}{|c|}{Fisher matrix} & $1.00^{+1.20}_{-1.20}$ & -- & $10.00^{+1.20}_{-1.20}$ & -- \\
  
  \hline

   \end{tabular}
\caption{ Recovered $68\%$ CI on the accretion parameter $f_{\rm Edd}$ 
and on the sky localization $\Delta\Omega$ for SOBHBs and for various realizations in the \emph{LISA+Earth} scenario. The last row gives the statistical error estimated with a Fisher-matrix analysis. The presence of accretion should be detected for super-Eddington accreting systems. The error on the sky position is always within the field of view of Athena and SKA, allowing (potentially) for electromagnetic followup.}\label{res_det}
  \end{table}}
  }

{
\renewcommand{\arraystretch}{1.5}
{\centering
  \begin{table}[th!] 
   \begin{tabular}{c|c|c|c|c|}
   \cline{2-5}
   
   & \multicolumn{2}{|c|}{$f_{\rm Edd}^{\rm injected}=0.1$}  & \multicolumn{2}{|c|}{$f_{\rm Edd}^{\rm injected}=1$}    
\\
   
   \cline{2-5}
   
   & $f_{\rm Edd}$ & $\Delta \Omega$ (${\rm deg}^2$) & $f_{\rm Edd}$ & $\Delta \Omega$ (${\rm deg}^2$)\\
   
   \hline
  
  \multicolumn{1}{|c|}{Light IMBHB}  & $0.14_{-0.14}^{+0.04}$  & $0.01$ &$1.05_{-0.11}^{+0.10}$ & $0.01$  \\

  \hline

 \multicolumn{1}{|c|}{Heavy IMBHB} & $0.17_{-0.17}^{+0.05}$ & $0.007$ & $1.04_{-0.27}^{+0.15}$ & $0.006$ \\

\hline

 \multicolumn{1}{|c|}{Fisher matrix} & $0.10_{-0.38}^{+0.38}$ & -- & $1.00_{-0.38}^{+0.38}$ & -- \\

\hline

   \end{tabular}
\caption{ Recovered $68\%$ CI on the accretion parameter $f_{\rm Edd}$ 
and on the sky localization $\Delta\Omega$ for the IMBHBs considered in this work, at redshift $ z=0.1 $. The statistical error estimated with our Fisher-matrix analysis is similar for the two IMBHBs. The presence of accretion should be detected for Eddington accreting systems. As for SOBHBs, the error on the sky position is always within the field of view of Athena and SKA. }\label{res_det_imbh}
  \end{table}}
  }

While overall in qualitative agreement, the differences between Fisher-matrix and MCMC results could be due to the effect of the priors, to the non-Gaussianity of the posterior distribution, to the treatment of the angles,
and/or to the finite SNR of the sources considered.
Nonetheless, the predicted errors on $f_{\rm Edd}$ are of the same order of magnitude in both treatments, confirming the main conclusions we drew for SOBHBs using the Fisher analysis.

\begin{figure*}[ht!]
	\centering
	        \includegraphics[width=0.8\textwidth]{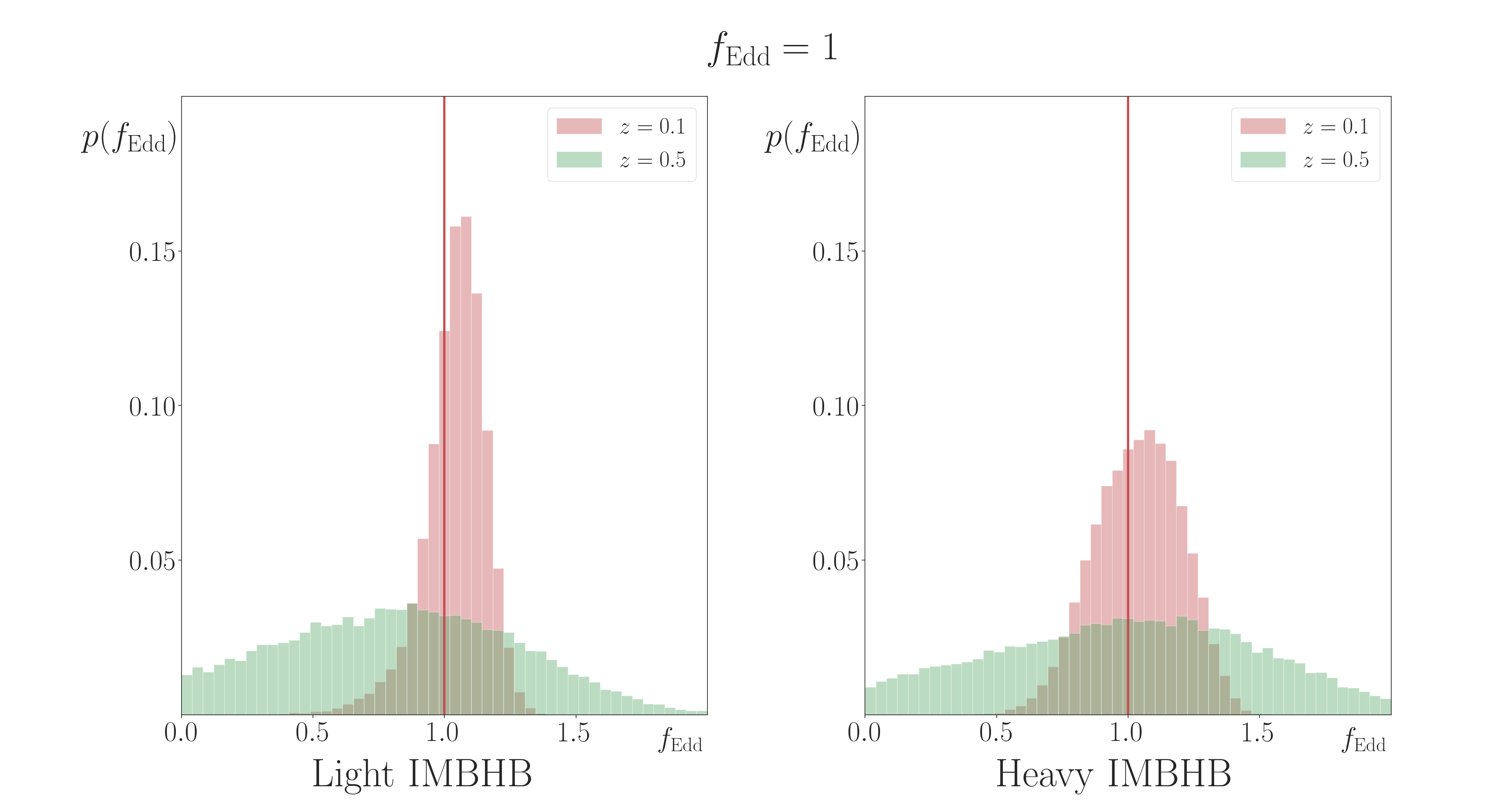}
	\caption{Marginalized distributions of $f_{\rm Edd}$ for our two IMBHB systems at redshifts $z=0.1$ (red) and $z=0.5$ (green) for an injected value of $f_{\rm Edd}=1$. Accretion can be measured in both systems at $ z=0.1 $, but not at higher redshift.}\label{fig:distr_imbh_reds}
\end{figure*}

In Fig.~\ref{fig:distr_imbh_reds} we compare how well can we recover $f_{\rm Edd}$ for IMBHBs at different redshifts, for injected $f_{\rm Edd}=1$. 
If the system is too far, the distribution tends to be flat and the effect of accretion is hardly noticeable.
% The peak in the distribution os not as clear as in figure Fig.~\ref{fig:distr1}. 
This is because of the lower SNR, but also because the detector-frame mass becomes larger at higher redshift, speeding up the evolution of the system and thus providing less information on negative PN-order modifications.

\begin{figure*}[ht!]
	\centering
	        \includegraphics[width=0.8\textwidth]{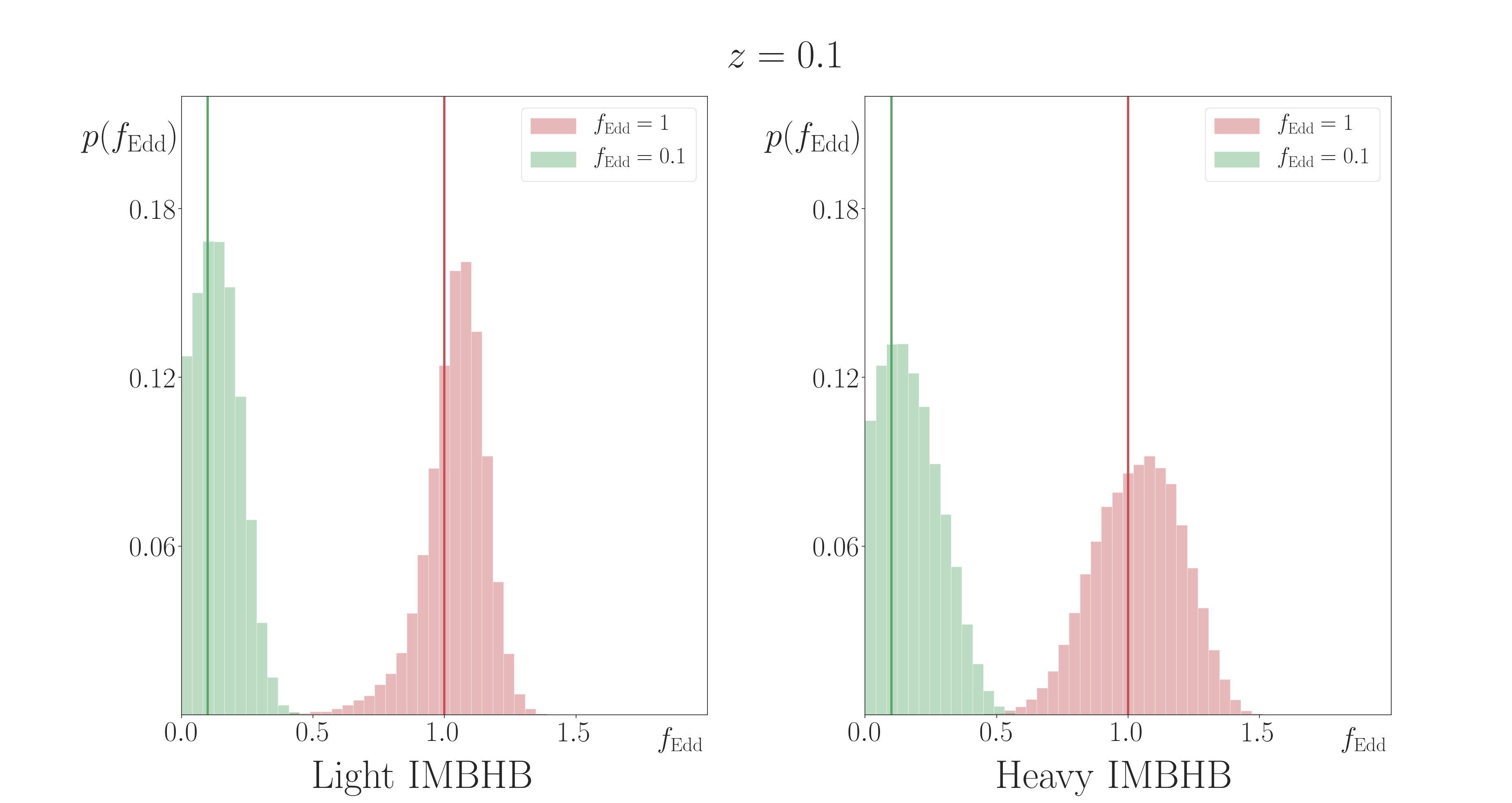}
	\caption{Marginalized distributions of $f_{\rm Edd}$ for our two IMBHB sources for injected values of $f_{\rm Edd}=0.1$ (green) and $f_{\rm Edd}=1$ (red) at $z=0.1$. Accretion at this redshift needs to be approximately Eddington-level or stronger to be measured.
	}\label{fig:distr_imbh_fedds}
\end{figure*}

Finally in Fig.~\ref{fig:distr_imbh_fedds} we show how well can we recover $f_{\rm Edd}$ in IMBHBs for an injected values of $f_{\rm Edd}=0.1$  at $z=0.1$.
As in the case of SOBHBs commented above, the marginalized distribution is compatible with $f_{\rm Edd}=0$, but the presence of a clear peak at $f_{\rm Edd}\neq0$ favours the presence of accretion.

%%%%%%%%%%%%%%%%%%%%%%%%%%%%%%%%%%%%%%%%%%%%
\subsubsection{Estimating biases}
%%%%%%%%%%%%%%%%%%%%%%%%%%%%%%%%%%%%%%%%%%%%
The above results indicate that if accretion is present it could lead to a measurable 
change in the GW signal.
Thus, if accretion is not taken into account, the estimation of other source parameters could be significantly biased. Since 
$f_{\rm Edd}$ correlates mostly with the intrinsic parameters of the source, the latter should be the most affected.

For SOBHBs in the \emph{LISA-only} scenario, we find that in the three cases (high, medium, and low SNR), the signal can be recovered by an effectual template with $f_{\rm Edd}=0$, i.e.,
we find a maximum for the posterior distribution which, in the worst cases, can be  incompatible with the injected real value. 
The SNR of this effectual template is very similar to 
the  injection's SNR (${\rm SNR}_{\rm inj}-{\rm SNR}_{\rm eff}\lesssim 0.7 $), and could thus 
trigger a detection. The bias in the parameter estimation and the relative drop in SNR is higher for lower SNR systems and for
higher injected accretion rates. The 
effectual template, in particular, has a higher chirp mass and a higher mass ratio, while the initial frequency is shifted towards higher 
values. In Figs.~\ref{fig:biases1} and~\ref{fig:biases10}
we show how this impacts the estimate of the masses and  time to coalescence 
for two representative values, $f_{\rm Edd}=1$ and $f_{\rm Edd}=10$. In both cases, we compare to the recovered distribution of masses for vacuum GR. 
\begin{figure*}[ht!]
	\centering
	        \includegraphics[width=0.8\textwidth]{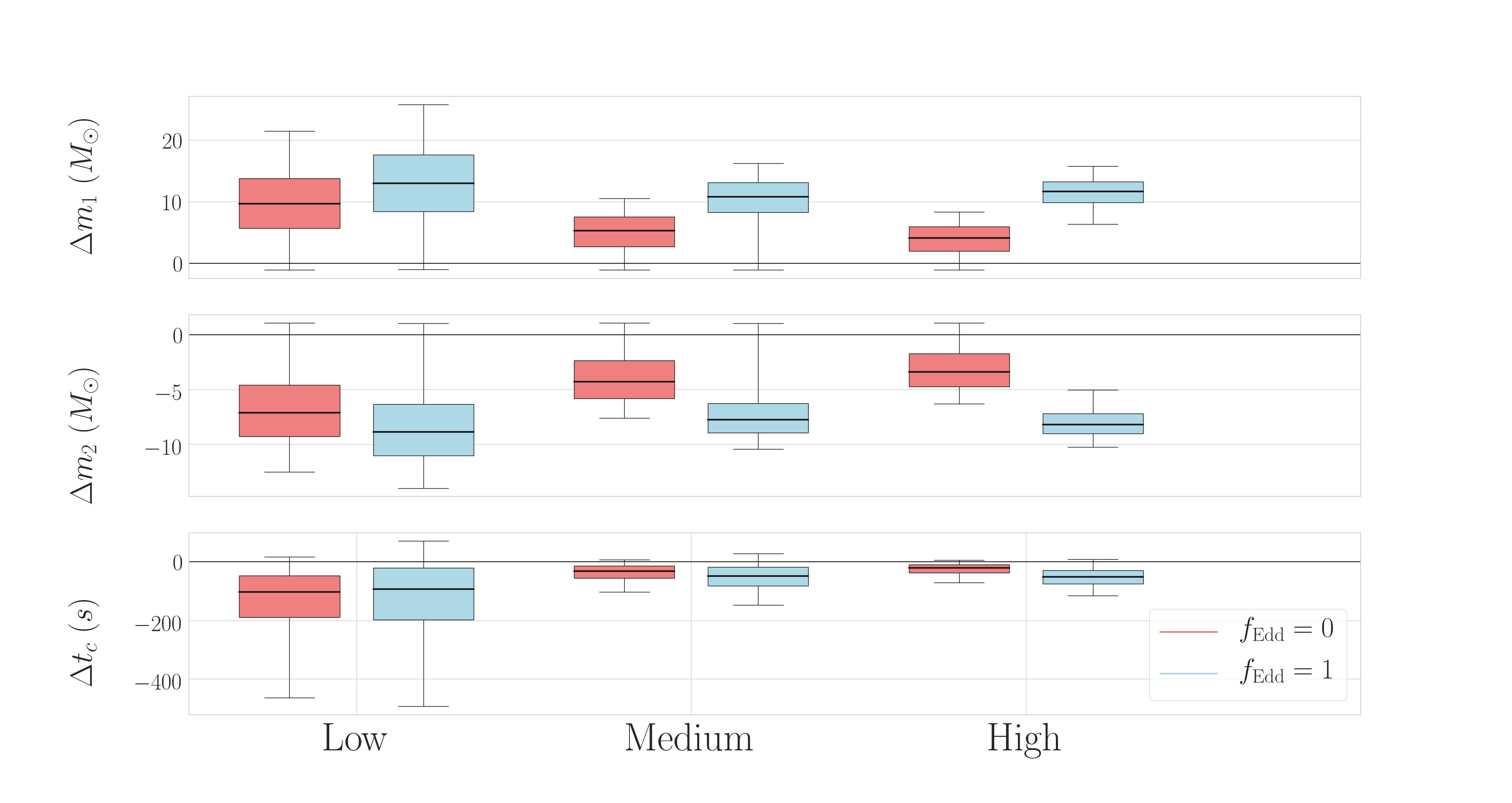}
    \centering
	\caption{Bias in the SOBH binary masses and time to coalescence induced by ignoring the corrections due to accretion when $f_{\rm Edd}=1$, for various angle
realizations in the \emph{LISA-only} scenario (blue) compared to the displacement found in vacuum systems. Boxes and whiskers delimit the $50\%$ CI and the $90\%$ CI, respectively, and both are centered around the median, indicated by lines inside the boxes. For this level of accretion bias is not significant, even for the high SNR realization. }\label{fig:biases1}
\end{figure*} 
\begin{figure*}[ht!]
	\centering
	        \includegraphics[width=0.8\textwidth]{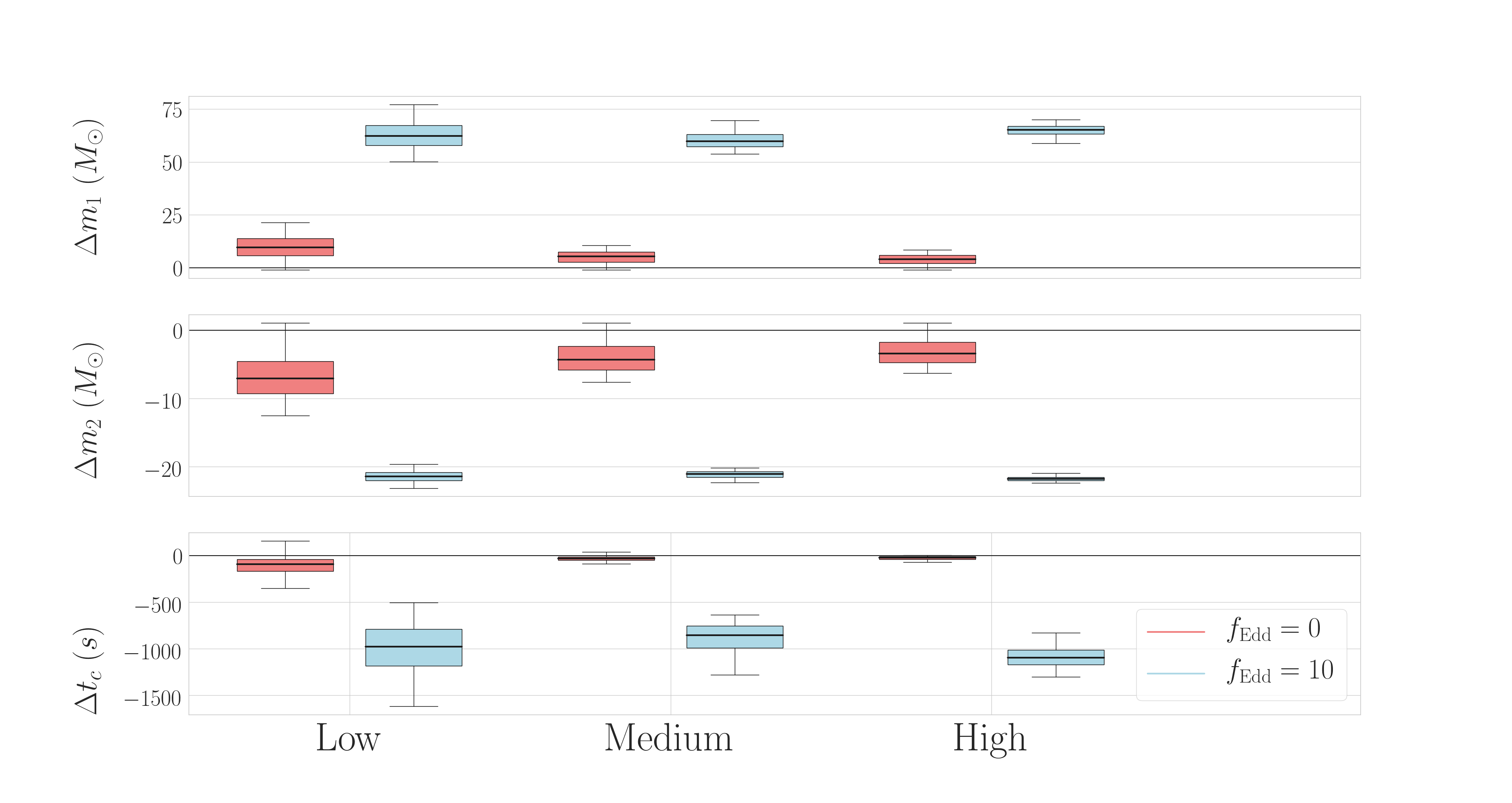}\\
    \centering
	\caption{Same as in Fig.~\ref{fig:biases1} but comparing $f_{\rm Edd}=0$ with $f_{\rm Edd}=10$. Bias here is significant for all SNR realizations. }\label{fig:biases10}
\end{figure*} 
The mass of the primary black hole is shifted toward higher values, whereas the secondary mass gets lower. As a result, the time to coalescence is underestimated. For super-Eddington accretion, this shift in time to coalescence is at the level of tens of seconds.
A multi-band observation could then help identify a bias due to accretion in the parameter estimation, 
since ground-based detectors would measure very precisely the time to coalescence when the signal enters in their 
band~\citep{Sesana:2016ljz}. 

\begin{figure*}[ht!]
	\centering
	        \includegraphics[width=0.8\textwidth]{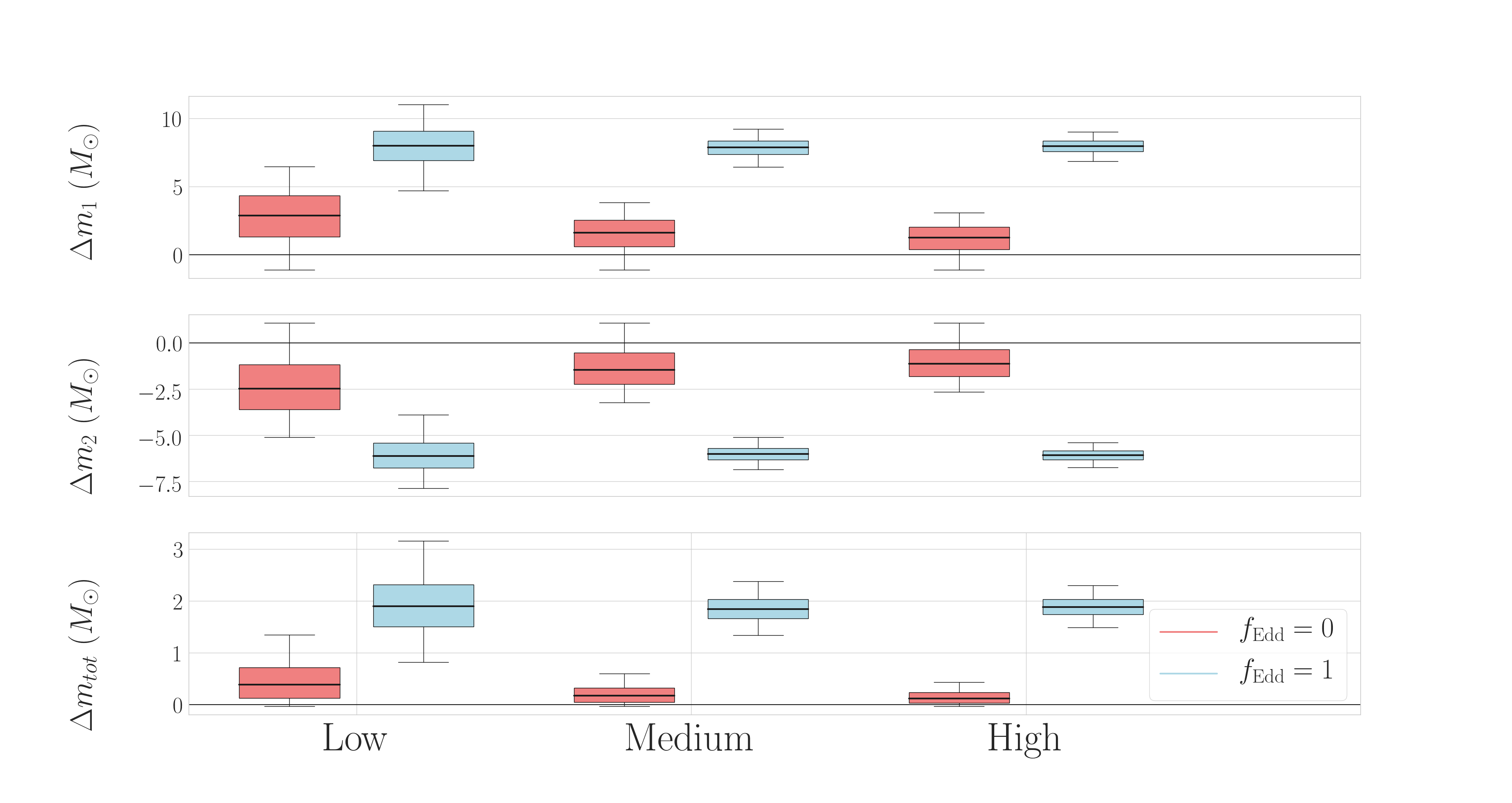}\\
    \centering
	\caption{Same as in Fig.~\ref{fig:biases1} but for the \emph{LISA+Earth} scenario. We show $m_{tot}$ rather of $t_c$, the latter being fixed by the narrow prior in this scenario. Bias in the masses can still be significant for medium and high SNR realizations, despite the constraint provided by ground based detectors.}\label{fig:biases1_mb}
\end{figure*}

In order to estimate this possibility, we repeat the above analysis in the \emph{LISA+Earth} scenario. In this case the time to coalescence is constrained to within $1\,{\rm ms}$ from its true value, so no bias in $t_c$ is possible. Nevertheless, signals can still be recovered by an effectual template, although with a larger mismatch from the true signal. 
In Fig.~\ref{fig:biases1_mb} we show the difference between the recovered masses and total mass and the injected values.
More in general, ground based detectors should be largely insensitive to these low frequency terms, as discussed in \cite{Carson:2019kkh}. In this forecast study, the projected constraint on -4PN terms with the planned third generation detector Cosmic Explorer is ten orders of magnitude worse than the projected constraint with LISA. We thus expect that observations with ground based detectors should not be biased by omitting -4PN terms. Therefore, for values of $ f_{\rm Edd} $ for which the LISA parameter estimation is significantly biased, the posterior distributions obtained with LISA and with ground based detectors might not even be compatible, which would hint at an unmodelled effect.

It is noteworthy that for the SOBHB events the sky localization is barely affected by accretion and remains excellent, as the distribution remains a Gaussian centered around the injected values with errors similar to the ones shown in Table~\ref{res_det}.
In the case of IMBHBs, on the other hand, there is also a bias in the sky localization, i.e. the injected value may lie outside the $90 \%$ CI. This is due to the very small errors in sky position, and in fact the the true localization is very close to the recovered one, within $0.05 \, {\rm deg}^2$. Therefore, for most realistic purposes the sky localization is satisfactorily recovered.

Since we did not consider any modification to the GW amplitude, there is no strong correlation 
between $f_{\rm Edd}$ and the luminosity distance $d_L$. Thus, when fixing $f_{\rm Edd}=0$ as we did here, there is no bias on the estimation of $d_L$, contrary to the Fisher-matrix analysis in \citet{Tamanini:2019usx}, who also used waveforms modifying GR at -4PN order in phase, but included the leading-order modification to the amplitude too.

%%%%%%%%%%%%%%%%%%%%%%%%%%%%%%%%%%%%%%%%%%%%%%%%
\subsection{Prospects for multiband and multimessenger astronomy} \label{sec:multi}
%%%%%%%%%%%%%%%%%%%%%%%%%%%%%%%%%%%%%%%%%%%%%%%%
According to our MCMC analysis, both SOBHBs and IMBHBs can be localized in the sky to within the fields of view of X-ray 
and radio instruments such as Athena WFI and SKA , $\Delta\Omega_{\rm Athena} = 0.4 \, {\rm deg}^2$, $ \Delta\Omega_{\rm 
SKA} = 0.5 \, {\rm deg}^2$ \citep{WinNT,2018CoSka..48..498M}.
This will allow the relevant region of the sky to be covered in a single viewing\footnote{In some cases, the correlation between the sky position angles can imprint an asymmetric shape to the localized region, which might therefore partially fall outside the field of view. However, this would still only require $ \mathcal{O}(1) $ viewings.}, thus potentially allowing for the coincident detection of an X-ray and/or radio counterpart to strongly-accreting black hole binaries.  
Even if the sky localization was biased, as might be the case for IMBHBs, we estimated that the true position would still fall inside the field of view of the instruments. In the following, we compute the X-ray and radio emission of the binaries, and estimate the necessary integration time for detection by a single instrument viewing.

We start by estimating the X-ray flux. For this purpose, we assume that the accretion process has radiative efficiency $ \eta=0.1$ (which is good approximation
at $f_{\rm Edd}<1$), and  that only a fraction $\eta_X=0.1$ of the EM radiation is emitted in X-rays (``bolometric correction''). We find the X-ray flux from a single accreting black hole to be
\begin{equation}\label{eq:fluxX}
F_X \simeq 1 \times 10^{-13} f_{\rm Edd} \, \Big(\frac{M}{ M_{\odot}}\Big)\Big(\frac{{\rm Mpc}}{d_L}\Big)^2 
%3 \times 10^{-17} f_{\rm Edd} \, \Big(\frac{M}{40M_{\odot}}\Big)\Big(\frac{415 {\rm Mpc}}{d_L}\Big)^2 
\, {\rm erg} \, {\rm cm}^{-2} \, {\rm s}^{-1} \,.
\end{equation}
This should be compared with the flux sensitivity of the Athena WFI for a given integration time, $T_{\rm int}$. 
Following~\citet{McGee:2018qwb}, Athena's flux sensitivity for a $5\sigma$ detection is
\begin{equation}
F_X^{\rm Athena}= 1 \times 10^{-15} \left( \frac{ 10^3 \, {\rm s}}{T_{\rm int}}\right)^{1/2} 
\, {\rm erg} \, {\rm cm}^{-2} \, {\rm s}^{-1}\, .
\end{equation}
The minimum integration time for a binary where only one black hole is emitting is then given by
\begin{equation}\label{mininttime}
T_{\rm int} \simeq 8  \times 10^{-2} \,  f_{\rm Edd}^{-2} \, \left( \frac{d_L}{ {\rm Mpc}} \right)^4 \, \left( \frac{ M_{\odot} }{M} \right)^2 
\, {\rm s} \, .
\end{equation}
Note that if the two black holes have similar mass and are both accreting, the cumulative flux 
is given by twice the value in Eq.~\eqref{eq:fluxX} and therefore the minimum integration time is one fourth of that in 
Eq.~\eqref{mininttime}. 

For the best-candidate SOBHB event in our synthetic astrophysical catalogues, the required exposure time is $T_{\rm int} \gtrsim  1 \times
10^6 \, f_{\rm Edd}^{-2} \, {\rm s}$. 
Thus, even if we were to assume $ f_{\rm Edd}\approx 1$, 
the integration time would have to be of several days. Assuming super-Eddington accretion $f_{\rm Edd}> 1$
is unlikely to help as the radiative efficiency is expected to be considerably lower than
our assumed $\eta=0.1$, i.e. the bolometric luminosity is not expected to significantly exceed the Eddington 
luminosity~\citep{Shakura:1972te,Poutanen:2006uc,2011arXiv1108.0396S}. 
Moreover, as previously discussed, high accretion rates in SOBHBs likely require environments with large gas densities, whose optical thickness further reduces the chances of an EM detection. 
For the considered IMBHB systems, the required integration time is between $ 24 $ and $ 2 $ hours for Eddington-level accretion, for the light and heavy systems, respectively. This estimate suggests that detection of X-ray counterparts will be possible for highly-accreting IMBHBs.

A binary system in external magnetic fields may also launch  dual radio jets,
which get amplified by the coalescence~\citep{Palenzuela:2010nf}
relative to similar jets observed in isolated black holes~\citep{Steiner_2012}.  
See also \citet{2012ApJ...749L..32M}  for simulations that yield $\sim$100 times larger (though less
collimated) fluxes than \citet{Palenzuela:2010nf}.
Assuming a fiducial value $ \eta =0.1 $ for the radiative 
efficiency of the process and $ \eta_{\rm radio} =0.1 $ for the fraction of emission in the radio band, the 
corresponding peak flux\footnote{The peak sensitivity is reached when the orbital velocity is equal to that of the innermost circular orbit.} is \citep{Palenzuela:2010nf,Tamanini:2016zlh}
\begin{multline}\label{key3}
F_{\rm flare} \simeq 2 \times 10^{-13} \, f_{\rm Edd} \,q^2 \, \left(\frac{D_{L}}{ {\rm Mpc}}\right)^{-2}  
\left(\frac{M}{ M_{\odot}}\right)
\\ \times 
{\rm erg} \, {\rm cm}^{-2} \,  {\rm s}^{-1}\,,
\end{multline}
where $q \le 1$ is the mass ratio. The flare flux can then be compared with the SKA-mid sensitivity in the phase 1 implementation.
The required sensitivity at frequency $ \nu_{\rm SKA} $ for SKA,
\begin{equation}\label{keyRADIO}
F_{\rm SKA} 
= 5 \times 10^{-16} \left(\frac{10^{-2} \, {\rm s}}{T_{\rm obs}}\right)^{1/2} \left(\frac{\nu_{\rm SKA}}{{\rm 
		GHz}}\right)  \, {\rm 
	erg} \, {\rm cm}^{-2} \,  {\rm s}^{-1}\,,
\end{equation}
is reached for an observation time $T_{\rm obs}\sim 10^{-2}\,{\rm s}$ for our best SOBHB event. The observation time should be smaller than the duration of the merger (i.e., the duration of the flare) for the system~\citep{Steiner_2012}, $T_{\rm flare} \sim 25 \,\, \frac{M}{100 M_{\odot}}$ ms. This condition is not satisfied for SOBHBs. There is however the concrete possibility to detect a signal in the radio band for IMBHBs, for which for the light and heavy systems $ T_{\rm obs} \approx 40 $ -- $  4 \ {\rm ms} < T_{\rm flare} $. The performance of full SKA should 
improve by an order of magnitude with respect to Eq.~\eqref{keyRADIO}, reducing the required integration time by a factor $ 100 $.
%%%%%%%%%%%%%%%%%%%%%%%%%%%%%%%%%%%%%%%%%%%%%%%%%%%%%%%%%%%%%%%%%%%%%%%%%
\section{Discussion}\label{sec:disc}
%%%%%%%%%%%%%%%%%%%%%%%%%%%%%%%%%%%%%%%%%%%%%%%%%%%%%%%%%%%%%%%%%%%%%%%%%
SOBHBs and IMBHBs provide the opportunity to measure the effect of accretion, which might affect the GW waveform at low 
frequencies. Our analysis suggests that a multiband detection with LISA and a ground-based detector will be able to 
measure the accretion parameter of strongly-accreting SOBHBs to within $50\%$ precision for a few events. For these 
systems, neglecting accretion in the waveform template might lead to biases in the recovered binary parameters. These biases can be alleviated by an accurate measurement of the time of coalescence by a ground-base detector.

IMBHBs in the local universe, if they exist as LISA sources, might also provide very accurate measurements of the accretion 
rate. Overall, for these systems the effect of accretion should be included in the waveform to avoid bias in the 
intrinsic binary parameters.

Finally, accretion does not affect sky localization by LISA for SOBHBs and it impacts that of 
IMBHBs only mildly. In both cases, the measurement errors are typically well within Athena's and SKA fields of view. 
Furthermore, the X-ray flux expected from 
strongly-accreting binaries is comparable with Athena's sensitivity and is well above the sensitivity of future 
missions such as Lynx~\citep{LynxTeam:2018usc}. Likewise, in the case of jets the radio signal from IMBHBs could be 
detectable by SKA. Our analysis shows that the simultaneous operation of Athena/SKA and LISA would 
therefore provide the thrilling opportunity to detect the EM counterpart of highly accreting black hole binaries.

%%%%%%%%%%%%%%%%%%%%%%%%%%%%%%%%%%%%%%%%%%%%%%%%%%%%%%%%%%%%%%%%%%%%%%
\begin{acknowledgments}
%%%%%%%%%%%%%%%%%%%%%%%%%%%%%%%%%%%%%%%%%%%%%%%%%%%%%%%%%%%%%%%%%%%%%%%
We are indebted to Davide Gerosa, Cole Miller, Luigi Stella and Alex Nielsen for insightful comments.
AC acknowledges support from the``Generalitat Valencian'' (Spain) through the "plan GenT" program (CIDEGENT/2018/019), as well as national grants FPA2014-57816-P, FPA2017-85985-P and  
the European projects H2020-MSCA-ITN-2015//674896-ELUSIVES.
LS acknowledges that research at Perimeter Institute is supported by the Government of Canada through Industry Canada 
and by the Province of Ontario through the Ministry of Research and Innovation.
EB acknowledges financial support provided under the European Union's H2020 ERC Consolidator Grant
``GRavity from Astrophysical to Microscopic Scales'' grant agreement no. GRAMS-815673.
PP acknowledges financial support provided under the European Union's H2020 ERC, Starting 
Grant agreement no.~DarkGRA--757480, and under the MIUR PRIN and FARE programmes.
This project has received funding (to EB and AT) from the European Union's Horizon 2020 research and 
innovation programme under the Marie Sklodowska-Curie grant agreement No 690904.
The authors would like to acknowledge networking support by the COST Action CA16104 and 
support from the Amaldi Research Center funded by the MIUR program ``Dipartimento di 
Eccellenza'' (CUP: B81I18001170001). 

\end{acknowledgments}
%%%%%%%%%%%%%%%%%%%%%%%%%%%%%%%%%%%%%%%%%%%%%%%%%%%%%%%%%%%%%%%%%%%%%%%%%%%%%%
\appendix
%%%%%%%%%%%%%%%%%%%%%%%%%%%%%%%%%%%%%%%%%%%%%%%%%%%%%%%%%%%%%%%%%%%%%%%%%%%%%%
\section{Accretion term in the GW waveform}\label{app:waveform}
%%%%%%%%%%%%%%%%%%%%%%%%%%%%%%%%%%%%%%%%%%%%%%%%%%%%%%%%%%%%%%%%%%%%%%%%%%%%%%
An accreting binary can be described by a  Hamiltonian $H(\boldsymbol{q},\boldsymbol{p})$, where the masses vary adiabatically.
As shown for instance in \citet{Landau} and \citet{Sivardire1988}, the action variables $I_q=\oint p {\rm d} q/(2 \pi)$ are adiabatic invariants. 
In our case, working in polar coordinates $r, \phi$ and in the center of mass frame,
we then have that $I_\phi=p_\phi$ and $I_r=\oint p_r {\rm d} r/(2 \pi)$ are conserved under accretion. The latter implies that
circular orbits remain circular under accretion, while the former is 
 equivalent to the conservation of the orbital angular
momentum  under accretion.

Then, to leading order, angular momentum is only lost through GWs~\citep{Peters:1964zz}, 
\begin{equation}\label{keybis}
\dot{L}_{GW} = -\frac{32 \, m_1^2 \, m_2^2 \,  M^{1/2}}{5 \, r^{7/2}} = -\frac{32}{5 } \mu^2 \omega ^{7/3} \, M^{4/3} \, .
\end{equation}
Defining the reduced angular momentum $ \tilde{L}_z=L_z / \mu M =\sqrt{r/M} $, the evolution of the binary can be obtained through
\begin{equation}\label{eq:equationGW}
\dot{\tilde{L}}_z = \frac{\dot{L}_{GW}}{\mu M}  \, .
\end{equation}
Integrating Eq.~\eqref{eq:equationGW}, we find the evolution of the orbital frequency,
\begin{equation}\label{eq:omegaGW}
\omega_{\rm GW}(t) = \left(\omega _0^{-8/3}-\frac{256  }{5 
} M^{2/3} \, \mu \, t\right)^{-3/8}\,,
\end{equation}
%%%
where $\omega_0=\pi f_0$ is the initial orbital frequency. The time as a function of the orbital frequency is found inverting this expression,
\begin{equation}\label{tGW}
 t_{\rm GW} (\omega) = t_c  -\frac{5}{256 \,  \mu \,  M^{2/3} \, \omega ^{8/3}} \, ,
 \end{equation}
 where $ t_c  $ is the merger time in the Newtonian approximation.
In the stationary-phase approximation, the GW phase reads~\citep{PhysRevD.49.2658,Maggiore}
\begin{equation}
\phi_{\rm GW}(f)= 2 \pi f t_c+ \phi_c - 2 \int_t^{t_c}\omega_{\rm GW}(t')dt' \nonumber = 2 \pi f t_c+\phi_c + \frac{3}{4} \left(8 \pi \mathcal{M} f \right)^{-5/3} \, ,
\end{equation}
where $ \phi_c $ is the phase at merger.

We shall now compare these known results with what happens in the presence of mass accretion. We assume that the binary is surrounded by gas and that both bodies are accreting mass at a same fraction of the Eddington rate,

\begin{equation}\label{mit}
m_i(t) = m_{i,0} \, e^{ f_{\rm Edd} \,  t  /\tau} \,,
\end{equation}
where $ \tau= 4.5\times 10^{7}\, {\rm yr}$ is known as the Salpeter time 
scale and $m_{i,0}$ is the initial mass of the $i$-th body.
When this time dependence is taken into account in the expression for the angular momentum, 
Eq.~\eqref{eq:equationGW} acquires an extra term:
\begin{equation}\label{eq:equationaccr}
\dot{\tilde{L}}_z = \frac{\dot{L}_{GW}}{\mu M} - \tilde{L}_z \frac{\dot{(\mu M)}}{\mu M} \, .
\end{equation}
In this equation, all masses should be considered time dependent, except the ones appearing in the angular momentum radiated by GWs. This is because accretion cannot be considered adiabatic compared to GW emission.\\
Accretion will in general be accompanied by a drag force $\vec{F}_{{\rm drag}}$ due to the fact that the accreted 
material carries some angular momentum. This effect can be quantified as
\begin{equation}
	\vec{F}_{{\rm drag},i}=\dot{m}_i(\vec{v}_{\rm gas}-\vec{v}_{i}) \, .
\end{equation} 
for each mass, where $\vec{v}_i$ is the velocity of the $i$-th body. 
For simplicity we parametrize this effect with a constant factor $\xi $, fixed by the relative velocity between the gas and the perturber~\citep{Barausse:2007dy,Barausse:2014tra}, 
\begin{equation}
\vec{F}_{{\rm drag},i}\simeq - \xi \, \dot{m}_i  \vec{v}_{i} \quad \rightarrow \quad \dot{L}_{\rm drag} = - \xi \, \dot{\mu} \,  r^2 \omega \, . 
\end{equation}
Note that the parameter $ \xi $ can be positive (drag) or negative~\citep[pull, see e.g.][]{Gruzinov:2019gpd}.
At leading order in $ f_{\rm Edd} \xi $, the term $ \dot{L}_{\rm drag}/\mu M $ should be added to the right-hand side of Eq.~\eqref{eq:equationaccr} to take the effect of the drag into account. 

We can now solve the total angular momentum variation equation for the orbital frequency,
\begin{equation}\label{omegaacc}
\omega_{\rm acc}(t) =
5^{3/8} e^{f_{\text{Edd}} \frac{(3 \xi +5) }{\tau }t} \left(5 \, \omega _0^{-8/3}-\frac{768 \, \mu _0 \, M_0^{2/3} \tau  \left(e^{f_{\text{Edd}} \frac{(24 \xi +35) }{3 \tau } t}-1\right)}{(24 \xi +35) f_{\text{Edd}}}\right)^{-3/8} ,
\end{equation}
where $M_0$ and $\mu_0$ are the initial values of the total and reduced mass, respectively.
This expression cannot be inverted exactly to find $t=t(\omega) $. We therefore use a perturbative expansion 
valid when the accretion correction is small, i.e. we assume $ t_{\rm acc}(\omega) = t_{\rm GW}(\omega) + 
f_{\rm Edd} \, t^{(1)}_{\rm acc}(\omega) + 
\mathcal{O}( f_{\rm 
Edd}^2)$. We verified this to be an excellent approximation in all realistic situations,
%. We stress that the accretion correction is small compared to the unperturbed case also, 
including when $f_{\rm Edd} \sim 1-100$. 
	This is because the dimensionless parameter always appears in the combination $f_{Edd}\cdot t/\tau$, which is always small for the evolution times scales that we consider. %That is to say, the expansion can be also thought as an expansion in $f_{Edd}/\tau$.
	%Notice we expand our expression in function of the dimensionless parameter $f_{Edd}$; however it always appears in the combination $f_{Edd}\cdot t/\tau$, which is small even for large $f_{Edd} \sim 10-100$. That is to say, the expansion can be also thought as an expansion in $f_{Edd}/\tau$

%In other words, 
%in all realistic cases the perturbative regime is adequate.}

In terms of the GW frequency, we find  
\begin{align}\label{t1acc}
t^{(1)}_{\rm acc}(f) = - \frac{25 \left(\pi ^{16/3} f^{16/3} (24 \, \xi +35)-3 \pi ^{16/3} f_0^{16/3} (8 \, \xi +15)+10 \, \pi ^{16/3} \left(f f_0\right){}^{8/3}\right)}{393216 \, \pi ^{32/3} \left(f f_0\right){}^{16/3} \mu _0^2 \, M_0^{4/3} \tau } .
%-\frac{125 \,  \left(f f_0\right){}^{2/3} \left(7 f^{16/3}+2 \left(f f_0\right){}^{8/3}-9 f_0^{16/3}\right) }{393216 \, \pi ^{16/3} \, f^6 f_0^6 \, \mu _0^2 \, M_0^{4/3} \, \tau }\,.
\end{align}
%%%
Finally, we can compute the contribution of accretion to the GW phase in the stationary phase approximation, $ h \sim |h | \, e^{i \phi} $, at first order in perturbation theory,  i.e. $ \phi \simeq \phi_{\rm GW} + \phi_{\rm acc}  =2 \pi f  \left( t_{\rm GW} +f_{\rm Edd}\, t^{(1)}_{\rm acc} \right)  - \int_0^{t_{\rm GW}+f_{\rm Edd} \,  t^{(1)}_{\rm acc}} 2 \, \omega_{\rm acc}  \, {\rm d} t $.  We find, again as a function of the GW frequency,
\begin{align}\label{phaseA}
\phi_{\rm acc} = & - f_{{\rm Edd}} \,  (8 \, \xi +15) \, \frac{ 75  \, \mathcal{M}_0 }{851968 \,  \tau } \, \left(\pi  f 
\mathcal{M}_0\right){}^{-13/3} 
+ f_{{\rm Edd}} \, \frac{25 }{32768 \, \pi ^{8/3}\, f_0^{8/3}\, \mathcal{M}_0^{5/3} \, \tau } \,  \left(\pi  f 
\mathcal{M}_0\right){}^{-5/3}
 \nonumber \\
&+ f_{{\rm Edd}} \, (3 \,  \xi + 4 ) \, \frac{25 }{19968 \,  \pi ^{13/3} \, f_0^{13/3} \,  \mathcal{M}_0^{10/3} \tau } - f_{{\rm Edd}} \, 
(24 \, \xi +35) \, \frac{25  }{196608 \, \pi ^{16/3} \, f_0^{16/3} \,  \mathcal{M}_0^{13/3} \tau } \, \left(\pi  f \mathcal{M}_0\right) \, 
.
\end{align}
In the expression above, the terms linear in frequency and independent of frequency can be reabsorbed in the definition of the time to coalescence $ t_c $ and the phase at coalescence $ \phi_c $, respectively. 
Eq.~\eqref{phaseA} tells us that the GW signal will be dominated by the effect of accretion if the frequency is sufficiently low.
By comparing the size of the leading order phase term (0PN) in the vacuum waveform and the -4PN term induced by accretion, we find that accretion is the dominant effect at frequencies below
\begin{equation}\label{key3bis}
f_{\rm acc} \simeq \frac{1}{\pi} \left(\frac{25}{3}\frac{45}{6656} \frac{f_{\rm Edd}}{\tau}\right)^{3/8} 
\mathcal{M}^{-5/8} \, .
\end{equation}
While in Eq.~\ref{phaseA} we show all the terms of the expansion, we have
verified that the -4PN term dominates. The inclusion of the 0PN term
changes the results of the main text by less than $1\% $. This is expected
since most of the binary evolution in the LISA band takes place at
large separation/low frequencies.
%
%\textbf{where we have restored the physical units for ease of comparison with Eq.~\eqref{key}.}
% 

In the analysis presented in the main text we discarded the terms proportional to the drag coefficient $\xi$, which would add an additional parameter in our waveform and require proper modeling of the distribution of the gas and its velocity around the black holes. From the functional form of Eq.~\eqref{phaseA} we can see that neglecting the drag does not affect the frequency dependence of the GW phase, while it might affect the size of the effect. However, $ f_{\rm Edd} $ and $ \xi $ enter the two leading terms in Eq.~\eqref{phaseA} in different combinations,
which would help disentangle the two effects.
Indeed we checked that for generic values of $\xi$ the time and phase shifts presented in Fig.~\ref{fig:dphiacc} do not vary dramatically.

\section{Details on the MCMC analysis}\label{app:MCMC}

Using Bayes' theorem we compute the posterior distribution for $\zeta$, the multidimensional vector parameterizing a 
waveform template $h$ given observed data $d$: 
\begin{equation} \label{bayes} 
 p(\zeta|d)=\frac{p(d|\zeta)p(\zeta)}{p(d)}
 \end{equation}

 For the prior, $p(\zeta)$, we assume a flat distribution in $m_1$ and $m_2$ with $m_1 \geq m_2 \geq 3 M_{\odot}$, 
flat in spin magnitude between $-1$ and $1$, volume uniform for the source localization and 
flat in the source orientation, its polarization and its initial phase. In the \emph{LISA-only} scenario we assume a flat prior in initial frequency and in the \emph{LISA+Earth} scenario we use instead a Gaussian prior centered around the true value of $t_c$ of width $\sigma_{t_c}=10^{-3} {\rm s}$. 
Assuming Gaussian noise, the likelihood is 
given 
by $p(d|\zeta)=e^{-\frac{1}{2}(d-h|d-h)}$ where parenthesis denote the inner product defined by:
$(h_1|h_2)=4 \mathfrak{Re} \left(\int \frac{\tilde{h_1}(f) \ \tilde{h_2}^*(f)}{S_n(f)} {\rm d}f \right)$.
In the denominator, $S_n(f)$ is the detector power spectral density, indicating the level of noise at a given frequency.
\\To sample the posterior distribution we use a Metropolis Hashtings Markov Chain Monte Carlo~(MHMCMC)~\citep{Karandikar2006,10.2307/2684568} 
algorithm that we designed for this problem. More details will be given in an upcoming publication~\citep{Marsat,Toubiana20_1}. The 
basic idea of the algorithm is to explore the parameter space through a Markov chain generated with a symmetric 
proposal 
$\pi$, $\pi(\zeta_1,\zeta_2)=\pi(\zeta_2,\zeta_1)$. Starting from a point $\zeta_0$, we accept the 
proposed 
point $\zeta_p$ with a probability given by the ratio of the posterior distribution, 
$\frac{p(\zeta_p)}{p(\zeta_0)}$. By 
doing so we accumulate samples representing the distribution. 
In order to increase the sampling efficiency, we parametrize the waveforms with parameters for which --~based on 
the PN expressions~\citep{Buonanno:2006ui,Buonanno:2009zt}~-- we believe the posterior distribution is 
simpler. We take 
$\zeta=(\mathcal{M},\mu/M,f_0,\chi_s,\chi_a,f_{\rm Edd},\phi,\sin(\theta),\psi,\phi_0,\cos(\iota),\log10_{d_L})$ in the \emph{LISA-only} scenario. In the \emph{LISA+Earth} scenario we use $t_c$ instead of $f_0$. 
Here $\chi_s$ is the symmetric combination of spins
\begin{equation}
 \chi_s=\frac{m_1 \chi_1+m_2 \chi_2}{m_1+m_2} \, ,
\end{equation}
while $\chi_a$ is the corresponding antisymmetric combination,
\begin{equation}
 \chi_a=\frac{m_1 \chi_1-m_2 \chi_2}{m_1+m_2} \, .
\end{equation}
For the proposal $\pi$, we use a Gaussian distribution based on Fisher matrix. 
To ensure we have independent samples we downsample the chain using the autocorrelation length.

To strenghten our confidence in our MHMCMC we cross-checked our results obtained with it to the ones obtained with Multinest, a public nested sampling algorithm~\citep{Feroz:2008xx,skilling2006}.

\bibliographystyle{apsr}
\bibliography{Ref}

\end{document}